\documentclass{svjour2}
\usepackage{float}
\usepackage{subfig}
\usepackage{amssymb}
\usepackage{amsmath}
\usepackage{graphicx}

\begin{document}

\title{Testing Born's Rule in Quantum Mechanics for Three Mutually Exclusive Events}

\author{Immo  S\"ollner \and Benjamin Gsch\"osser \and Patrick Mai \and Benedikt Pressl \and Zolt\'an V\"or\"os \and Gregor Weihs}

\institute{Immo  S\"ollner$^*$, Benjamin Gsch\"osser, Patrick Mai, Benedikt Pressl, Zolt\'an V\"or\"os, and Gregor Weihs \at
Institut f\"ur Experimentalphysik, Universit\"at Innsbruck, Technikerstra{\ss}e 25, 6020 Innsbruck, Austria\\
$^*$ now at: Niels Bohr Institute, University of Copenhagen, Blegdamsvej 17, 2100 Copenhagen, Denmark}

\date{Received: March 6, 2011  / Accepted: July 26, 2011}

\maketitle

\begin{abstract}
We present a new experimental approach using a three-path interferometer and find a tighter empirical upper bound on possible violations of Born's Rule. A deviation from Born's rule would result in multi-order interference. Among the potential systematic errors that could lead to an apparent violation we specifically study the nonlinear response of our detectors and present ways to calibrate this error in order to obtain an even better bound.
\end{abstract}

\section{Theory}
\subsection{Introduction}
Two path interference is often regarded the characteristic phenomenon of any wave theory. It was first investigated by Thomas Young in 1801 when he tried to answer the question whether light was made of waves or particles. A similar experiment using electrons was repeated by Claus J\"{o}nsson in 1961 to show that the interference of two possible paths exists for electrons, as predicted by quantum mechanics. This is still often referred to as one of the most beautiful physics experiments of all times \cite{Rodgers02}. Born's rule allows us to calculate probabilities from quantum mechanical wavefunctions and results in interference terms that always originate from pairings of paths, but never triples, etc. Of course, multi-path interference will result in more complex interference patterns, but only by virtue of the multiple path-pair interference terms. Only very recently have there been investigations, experimental and theoretical, of possible new, higher-order phenomena going beyond the standard description of interference, when three or more paths are made to interfere \cite{Sinha09,Sinha10,Ududec10}. These investigations have been sparked by the seminal work of R. D. Sorkin \cite{Sorkin94}.

Out of an idea to implement quantum theory as measure theory on spacetime Sorkin analyzes interference as a deviation from the classical additivity (sum rule) of the probabilities of mutually exclusive events. If one considers more than two mutually exclusive events or paths $A$, $B$, $C$, $\ldots$ one can define a hierarchy of higher-order interference terms.

\begin{align}
I^{(0)}_{A} &= P_{A} \nonumber \\
I^{(1)}_{AB} &= P_{AB}-P_{A}-P_{B} \label{sumrules} \\
I^{(2)}_{ABC} &= P_{ABC}-P_{AB}-P_{AC}-P_{BC}+P_{A}+P_ {B}+P_{C} \nonumber
\end{align}

These three terms are the zeroth, first, and second order interference terms. If in a general theory the interference term $I^{(n)}$ is nonzero, this is indicative of $n^\mathrm{th}$ order (or $(n+1)$ path) interference and the theory is said to violate the $n^\mathrm{th}$ sum rule $I^{(n)}=0$. Here $P_A$ is the detection probability of an interference experiment with paths $A$, $B$, and $C$ when only path $A$ is open, $P_{AB}$ when paths $A$ and $B$ are open, and so forth.

If the zeroth sum rule is fulfilled, the measurement becomes trivial, and we will therefore only concern ourself with the case where $I^{(0)}_A\neq0$. The next higher sum rule, $I^{(1)}_{AB}=0$, holds for classical particles and is violated by quantum mechanics and classical waves. The violation $I^{(1)}_{AB}\neq0$ means that our system exhibits two-path interference. In classical wave theory and in quantum mechanics the second sum rule holds, $I^{(2)}_{ABC}=0$. In fact this is true for any theory where there is a square law relation between the energy (or probability) density and the field amplitude \cite{Sinha10}. This square law relation is of course Born's rule and is a basic postulate of quantum mechanics.

\begin{figure}
  \centerline{\includegraphics[width=\textwidth]{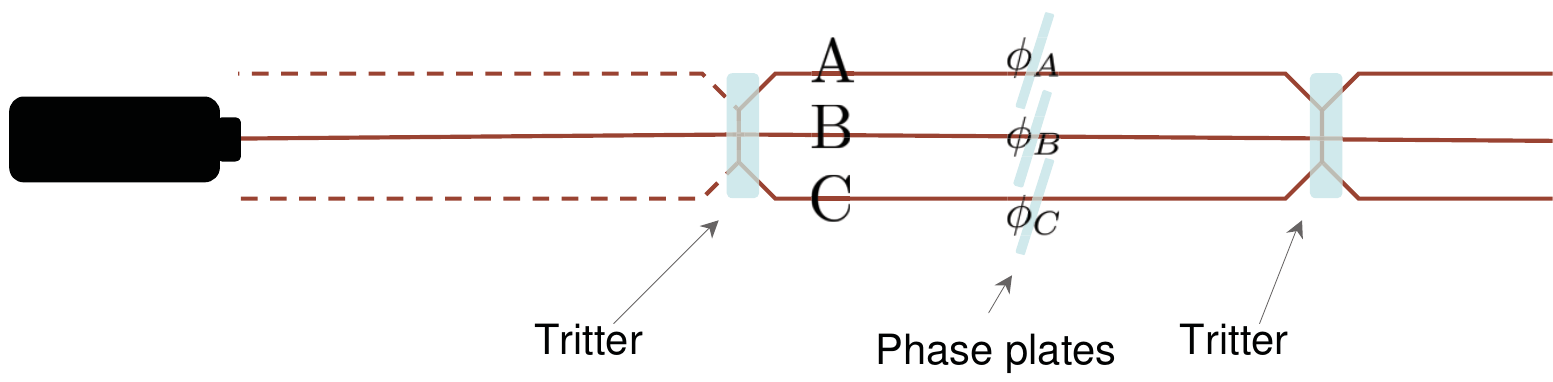}}
  \caption{Generic three-path interferometer composed of two three-way beam-splitters (also called 6-ports or ``tritters'' \cite{Mattle95a}) and a phase shift associated with each of the paths $A$, $B$, and $C$. Since only the relative phase matters, we will set $\phi_B=0$ without loss of generality.}
  \label{interferometer}
\end{figure}

The purpose of our experiment is a direct test of Born's rule by looking for any violation of the second sum rule. The mutually exclusive events are passages through three possible paths of an interferometer as shown in Fig.~\ref{interferometer}. In previous versions \cite{Sinha09,Sinha10} these events were approximated by passage of single photons through either of three adjacent slits in a diaphragm followed by detection in the far field. It has been pointed out, for example in Ref.~\cite{DeRaedt11}, that slits, in particular, if they are closely spaced are not entirely independent. This crosstalk will, however generally, vanish very rapidly with the separation of the respective paths. Therefore in an interferometer with a macroscopic path separation of several ten thousand optical wavelengths, the crosstalk caused by opening or closing another optical path will be much much smaller than what can be detected.

The experimental quantities that are being measured are the count rates $R_x$, where $x$ designates the open interferometer paths. The eight possible combinations of open and closed paths are then combined as in $I^{(2)}_{ABC}$ to form the quantity
\begin{equation}
  \epsilon =R_{ABC}-R_{AB}-R_{AC}-R_{BC}+R_{A}+R_{B}+R_{C}-R_{0},
\end{equation}
where $R_{0}$ is the background rate, taken with all interferometer paths closed. This is the experimental equivalent of Eq.~\ref{sumrules} and should therefore always be zero in quantum mechanics. Since we can not measure the probabilities in Eq.~\ref{sumrules} directly, the equations are written in terms of count rates, $R \propto P$. The last term in $\epsilon$ has to be subtracted to eliminate the total contribution from any background rate that is constant over all eight combinations, for example the detector dark count rate. We verified that the detector dark counts are indeed constant in time and independent of the history, i.e. of signals applied earlier, within our measurement precision. Beyond that it is difficult to check whether the background -- e.g. stray light -- is indeed constant. Any background caused by a closed shutter cancels identically by virtue of the definition of $\epsilon$. Likewise, imperfect coherence of the paths has no bearing on the result. Background contributions that depend in a non-trivial way on the positions of all the shutters would cause a non-zero $\epsilon$ and are thus sources of systematic errors that can only be eliminated by improvements of the experiment.

The quantity $\epsilon$ is further normalized by
\begin{equation}
    \delta=|R_{AB}-R_{A}-R_{B}+R_0|+|R_{AC}-R_{A}-R_{C}+R_0|+|R_{BC}-R_{B}-R_{C}+R_0|,
\end{equation}
which is the sum of the absolute values of the expected two-path interference terms. This newly defined quantity \cite{Sinha10}, $\kappa={\epsilon \over \delta}$, makes it more meaningful to compare the results from different experimental approaches, because it normalizes the hypothetical second order interference by the expected first order interference.

In the following we will discuss a new experimental approach to this measurement, a three-path interferometer, and will focus on one of the major reasons for experimental deviation from the expected null result, the nonlinearity of our detector. Using a simple model for the nonlinear behavior of the Avalanche Photodiode (APD) illuminated by a Poissonian light source, the deviation from the null measurement can be predicted, compared to experimental results, and finally used to correct this systematic error.

\subsection{Detector Nonlinearity}
The main reason for the nonlinearity of our detector is its dead-time. Each detection event of an APD is followed by a period of time during which the detector is unresponsive. During this time the detector can essentially be viewed as being turned off. If the efficiency of the detector during the active time periods, can be viewed as being independent of factors such as the the average count rate or the total incident rate then it can be ignored. Therefore the efficiency was set to one for the following discussion. To quantify this effect we want to describe the number of photons detected, $R^\mathrm{det}$, as a function of the total rate $R$ of incident photons per unit of time. The simple model presented below shows that the number of incident photons lost because their arrival time is within the dead-time caused by a previous detection does not grow linearly with the total number of incident photons, and therefore leads to a nonlinear measurement response of the detector.\cite{Davidson68}

An attenuated laser is used as the photon source in this experiment and the photon emission statistics can therefore be described by a Poissonian distribution so that the average number of incident photons is simply $RT$.\footnote{This holds for any stationary light source if the time interval is larger than the source's correlation times.} In Eq.~\ref{photonslost} it is shown that the rate of detected photons per second is equivalent to the rate of incident photons minus the number of photons that are expected to arrive within the dead time $\tau$ after each detection.
\begin{equation}\label{photonslost}
R^\mathrm{det}=R -  R^\mathrm{det}R\tau\end{equation}
Rearranging the equation to solve for $R^\mathrm{det}$ and $R$ respectively yields.
\begin{equation}\label{detectedphotons}
R^\mathrm{det}={R \over 1+R \tau}\ ,\ \ \ \ \ R={R^\mathrm{det} \over 1-R^\mathrm{det} \tau}
\end{equation}
So far we assumed that a detection can only be triggered when a photon is incident on the detector. For this model to be more accurate in the low intensity regime the effect of detector dark counts and other sources of constant background detections have to be taken into account. Consequently we can define the detector transfer function $f$ and its inverse $\bar f$ as
\begin{align}\label{detectedphotonswithdark}
    R^\mathrm{det}=f(R,R_0) &:= {R + R_0 \over 1+(R+ R_0) \tau} \\
    R=\bar f(R^\mathrm{det},R_0) &:= {R_0-R^\mathrm{det}(1+R_0\tau) \over -1+R^\mathrm{det} \tau}.
\end{align}
This model can now be used to predict an expected deviation from the zero measurement solely due to the nonlinear detector response. For this, however, it will be important to precisely determine the detector dead time and background count rate, and to verify that the model does describe the detector response.

\subsection{Calculating $\kappa^\mathrm{det}$}
In this section we show how the equations describing the detector nonlinearity, introduced above, are used to predict expected deviations from $\kappa=0$. To do this we calculate the incident intensities of the multi-path combinations from the measured values of the single-path cases $R_A^\mathrm{det}$, $R_B^\mathrm{det}$, and $R_C^\mathrm{det}$. Then the two and three-path incident rates are
\begin{align}
  R_{AB}  &= R_A + R_B + 2 \sqrt{R_A R_B} \cos{(\phi_A)} \\
  R_{ABC} &= R_A + R_B + R_C + 2 \sqrt{R_A R_B} \cos{(\phi_A)} \nonumber \\
          &+2 \sqrt{R_A R_C} \cos{(\phi_A-\phi_C)}+2 \sqrt{R_B R_C} \cos{(-\phi_C)},
\end{align}
where $\phi_A$ and $\phi_C$ are the phase shifts applied to paths $A$ and $C$, respectively and path $B$ experiences no phase shift, i.e. serves as the phase reference. The incident photon rates are calculated using the inverse detector transfer function, i. e. $R_{A}=\bar f(R_A^\mathrm{det},R_0)$, and similarly for $R_{B}$ and $R_{C}$.

From these inferred incident rates we can predict the rates $R_{AB}^\mathrm{det}$, $R_{AC}^\mathrm{det}$, $R_{BC}^\mathrm{det}$, and $R_{ABC}^\mathrm{det}$, and thus $\kappa^\mathrm{det}$, taking into consideration the nonlinear response of the detectors and background detections. Here only the detector dead time $\tau$, the background count rate $R_0$, and the measured rates $R_A^\mathrm{det}$, $R_B^\mathrm{det}$, and $R_C^\mathrm{det}$ are used as input parameters. We can then compare these to the values that we obtain directly from the data of all eight path combinations to see if there is any amount of second order interference beyond what can be explained by the detector nonlinearity.

Finally we could deduct the systematic error caused by the nonlinearity from the measured $\kappa$ value. Any such correction has to be done with caution because in general we may not be able to distinguish the nonlinearity of a detector from a nonlinearity of nature, i.e. a violation of Born's rule. A correction is only possible if the detector nonlinearity is being calibrated by a method that does not use interference, for example the beam combination method, which is detailed in the following section.

\section{Experiment}
\subsection{Testing detector nonlinearity}

\begin{figure}[htb]
    \begin{center}
    	\includegraphics[width=7.5cm]{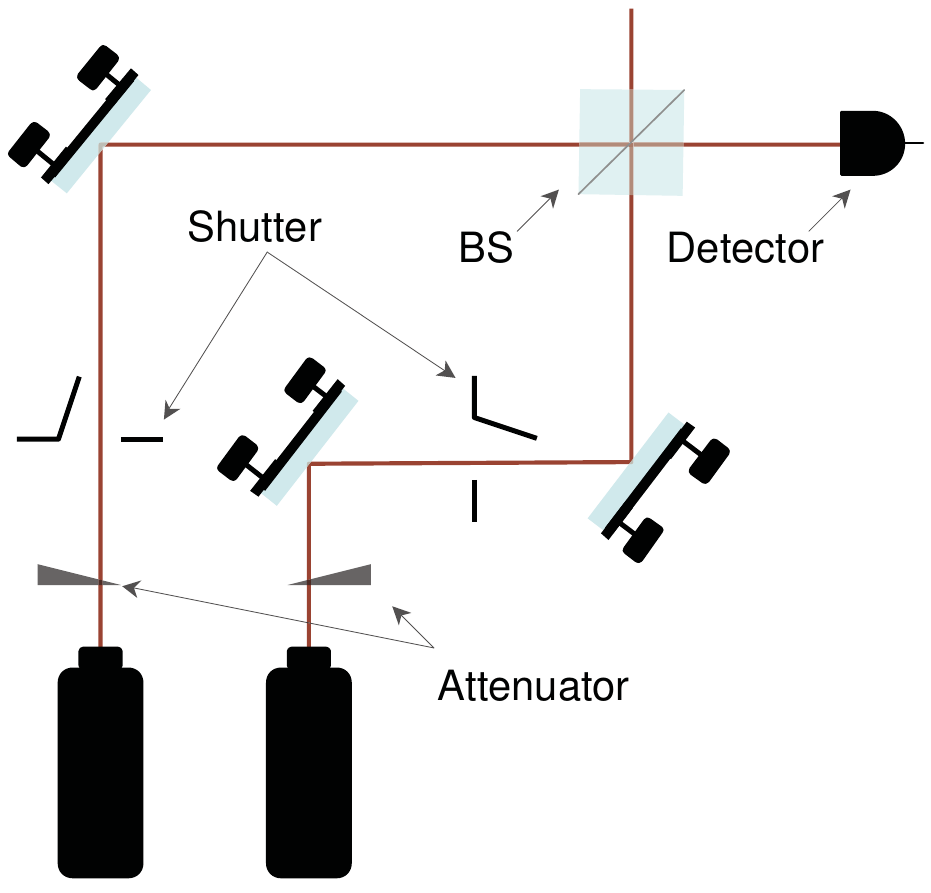}
    \end{center}
    \caption[Nonlinearity Setup]{The experimental setup being used for the characterization of the detector nonlinearity using the beam combination method.}
    \label{nonlinsetup}
\end{figure}

For a precise determination of the detector nonlinearity, i.e. the dead time $\tau$, we set up a direct measurement using the beam combination method. This method is based on the addition of non-interfering beams on the detector \cite{Yang94}. The setup in Fig.~\ref{nonlinsetup} was used for these measurements. For each individual measurement two independent laser beams were adjusted in their intensity and we measured four values, the dark counts, the count rate for each beam separately open and the total count rate when both beams were open. This is repeated for a variety of laser intensities in order to be able to map out the behavior of the detector for the desired range with increasing accuracy. The intensities are chosen by changing the attenuation in the two independent arms using variable attenuators. In order to infer the dead time, several assumptions had to be made, yet, fewer than for any other method we are aware of. Most importantly we assumed:
\begin{itemize}
\item That the intensities in the two arms will remain constant on the time scale it takes to record the four values, for this we have to examine the laser fluctuations and fluctuations on transmission of the attenuators.
\item That the dead time has a constant value for the range of incident intensities investigated.
\item That the detector efficiency, when it is active, is independent of factors like the total incident intensity and the average count rate.
\end{itemize}
The value of the dead time can be extracted from these measurements using a method similar to that presented in Ref.~\cite{Coslovi80}. A more detailed paper on the nonlinearity measurement and parameter estimation is under preparation.

The avalanche photon detector used in this experiment are specified to have a dead time of $\tau\lesssim50~\textnormal{ns}$, and our measurements lead to a dead time of $\tau = 47(2)\, \textnormal{ns}$ and a dark count rate of $ 284(4)\, \textnormal{s}^{-1}$.

In this nonlinearity measurement we use the incoherent addition of beams to measure the detectors response to an increased photon (or energy) flux, independent of how these rates are connected to the respective amplitudes. Therefore we can use this value to correct the systematic error in $\kappa$ to reveal any actual violation of Born's rule. In order to make this more convincing, we can always vary the overall rate and then check whether our estimated systematic error varies accordingly. Any deviation from the expected must then either be another systematic error or a violation of Born's rule.

\subsection{Three-path setup}
\begin{figure}[b]
\begin{center}
	\includegraphics[width=10cm]{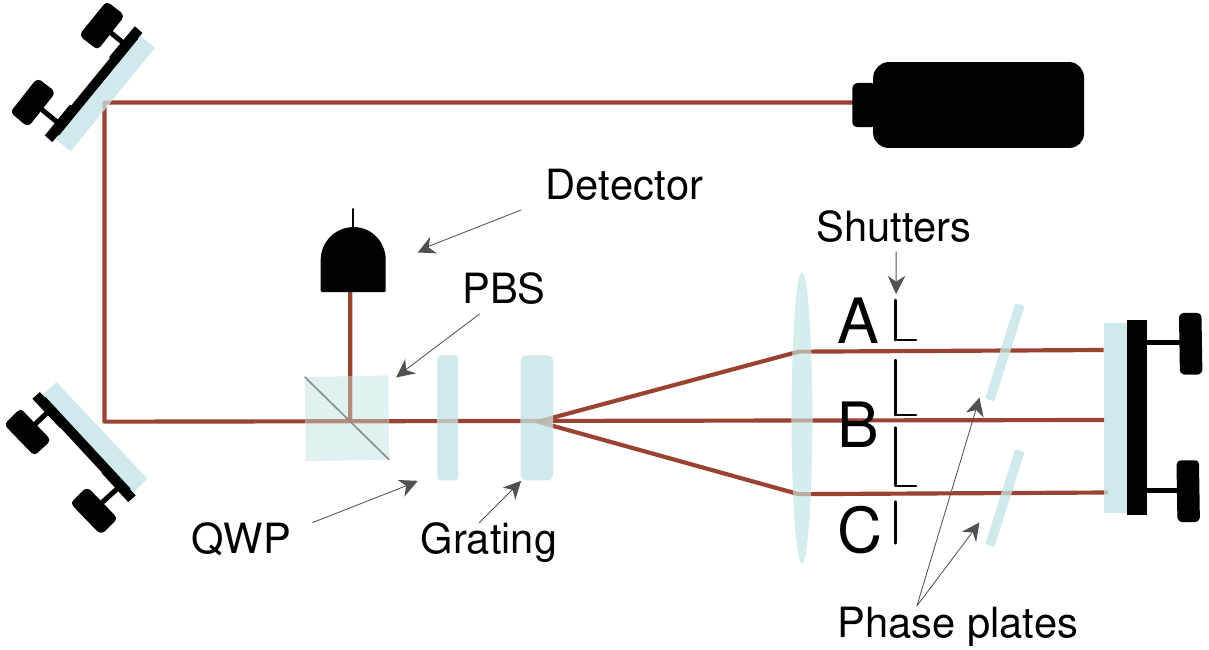}
	\caption[Nonlinearity Setup]{The experimental setup with the individual shutters for the three paths, and the two phase plates to set the position in phase space.}\label{threepathsetup}
\end{center}
\end{figure}

We built a three path interferometer using the zeroth and two first orders of a transmission grating as the three independent paths. Two of the paths pass through independently controlled phase plates, and all three paths are back reflected on themselves by a common mirror. The outgoing beam is separated from the incoming beam by a double pass through a quarter wave plate (QWP) and a polarizing beam-splitter, see Fig.~\ref{threepathsetup}.

The phase plates are $0.9\ \textnormal{mm}$ thick antireflection coated glass plates, mounted on motorized rotation stages. To compare the measured data to the modeled expectations, the rotation angles of the phase plates have to be transformed to the actual phases denoted by $\phi_A$ and $\phi_C$, where $n_1$ and $n_2$ stand for the refractive indices for air and glass respectively.

\begin{equation}
      \phi_{A}= {2 \pi \over \lambda} 2 d \left[ n_1 - n_2 +
    {n_2 - n_1 \cos\left(\theta_{A} - \theta'_A\right) \over
     \cos \theta'_A}\right],
 \end{equation}
where $n_2\sin\theta'_A=n_1\sin\theta_A$. The outgoing beam is spatially filtered with a pinhole to minimize the contributions from background caused by scattering on any of the surfaces in the setup and to allow us to measure a well-defined position in the space given by the two phases. Finally the beam is coupled into the detector via a multimode fiber.

A full measurement cycle consist of a 2-dimensional raster scan, using the phase plates shown in Fig.~\ref{threepathsetup}, with several measurements of $\kappa$ at every position in phase space. One $\kappa$ measurement consists of the eight possible combinations of three shutters independently open and closed. The sequence in which these eight combinations are taken for each $\kappa$, are randomized to avoid any effects caused by slow drifts of the laser. In addition to the complete phase space scan we chose a few particular phase settings, where we  performed a larger number of $\kappa$ measurements in order to reduce the statistical error.

\subsection{Results}
First, we measured the three different two-path interference visibilities and compared them to the achievable maximum values to check the quality of alignment. Here it is important to note that path $B$, the central path, has more power than either of paths $A$ and $C$ therefore we couldn't achieve high visibilities for the $AB$ and $AC$ path combinations and neither could we achieve 100\% visibility for the full three-path interferometer \cite{Weihs96a}. Then the three path interference pattern is measured and compared to the theoretical expectations.

\begin{figure}[H]
\begin{center}
 \subfloat[]{\label{theorythreepath}\includegraphics[width=0.48\textwidth]{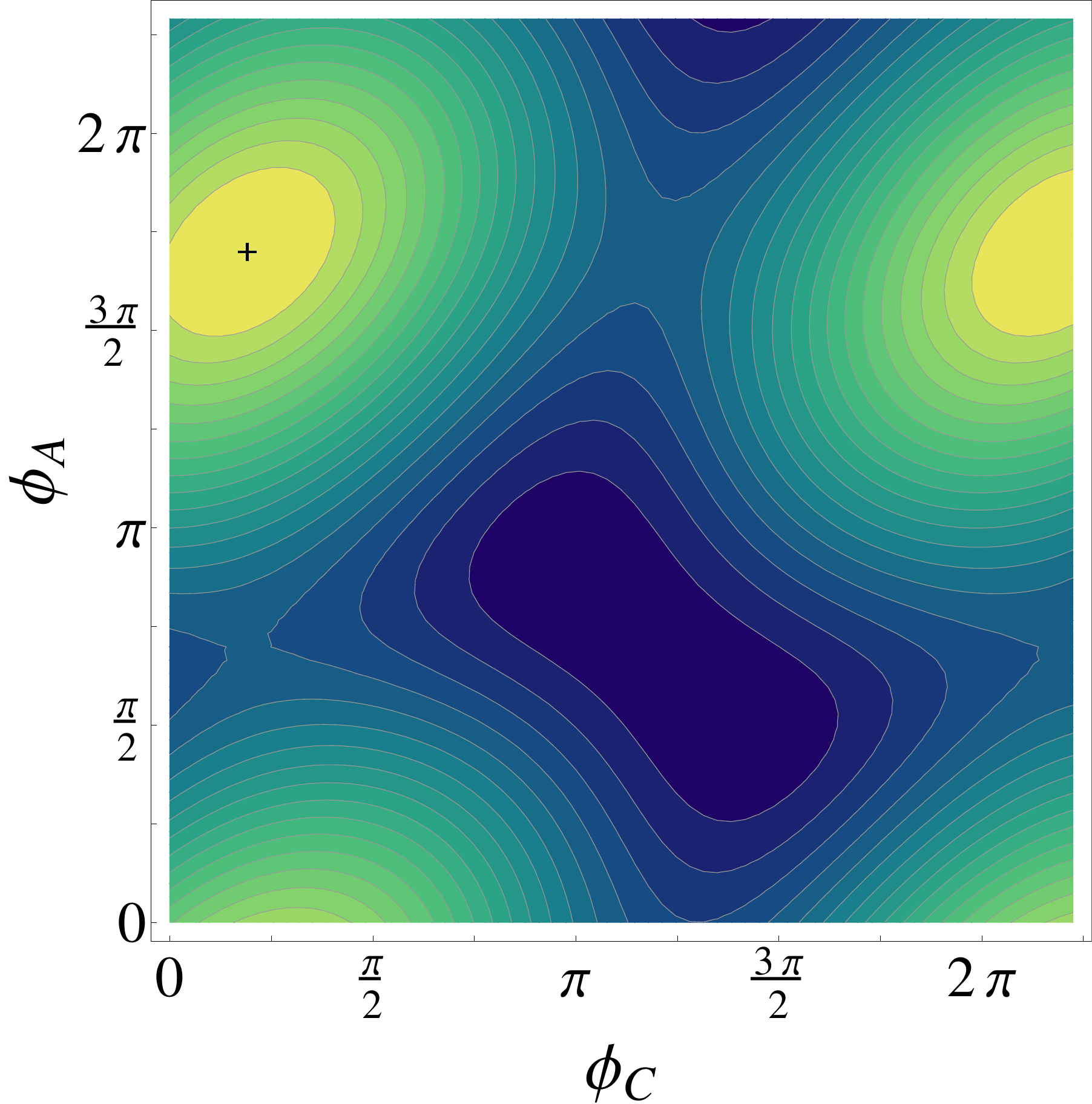}}  \ \ \
 \subfloat[]{\label{measuredthreepath}\includegraphics[width=0.48\textwidth]{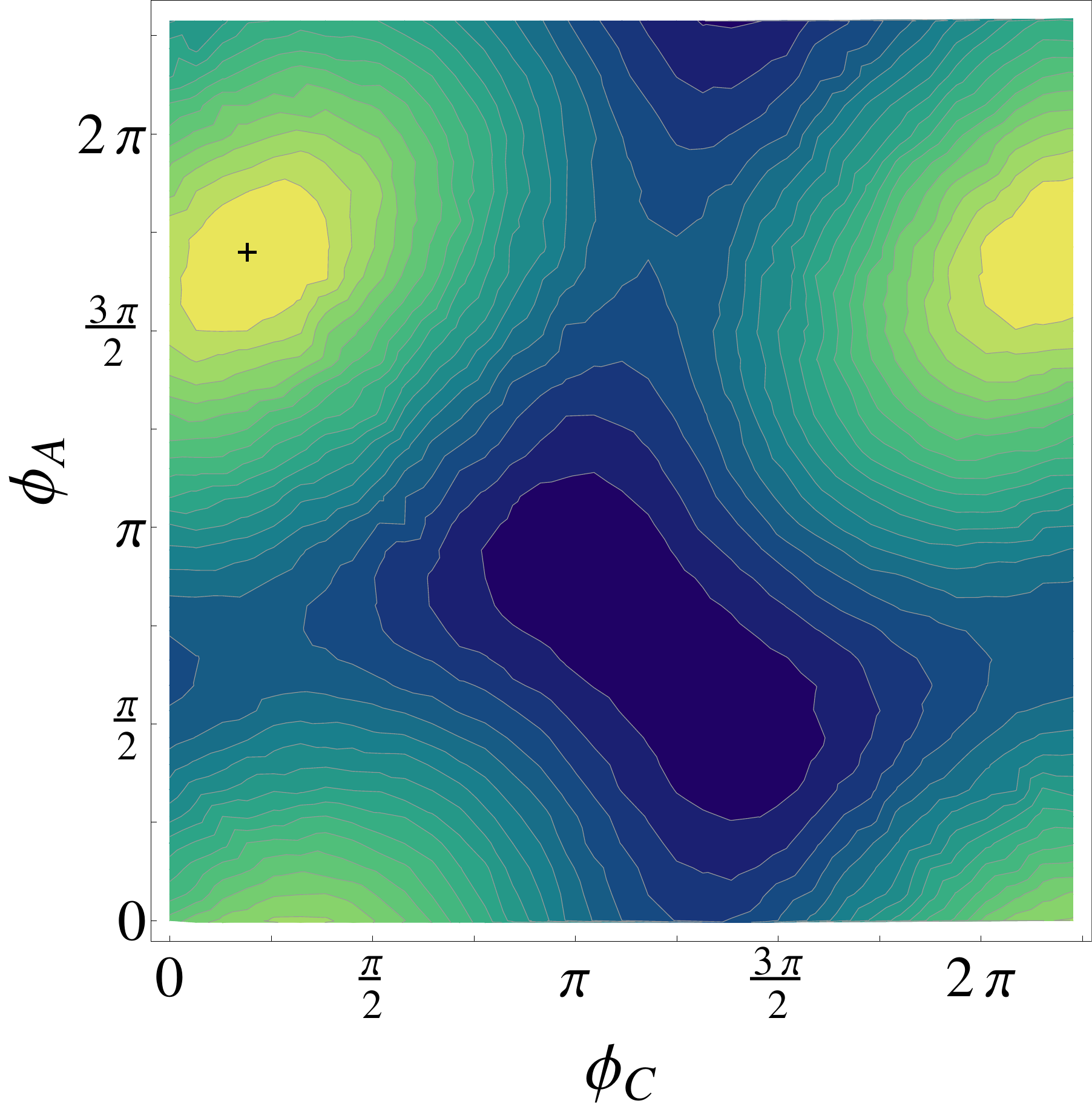}}
\end{center}
\caption{
Contour plots of the three-path interference pattern in two-dimensional phase space, where the values on the axes give the phase set by the two phaseplates. The crosses mark the phase space maximum used for more extensive $\kappa$ measurements. In (a) we show the theoretical model using the independent single-path intensities as input parameters. In (b) are the the measurement results at the same single-path intensities, $R_A^\mathrm{det}=2.08~\textnormal{kcps}$, $R_B^\mathrm{det}=5.76~\textnormal{kcps}$, $R_C^\mathrm{det}=1.99~\textnormal{kcps}$. The only variable used to enhance agreement between the two plots is a slight adjustment of the absolute starting position of the phase plates entered into the theoretical model. This was less then $1^{\circ}$ and had to be done since the initial starting position could not be determined with high enough accuracy.  }
\label{threepathdata}
\end{figure}

\begin{figure}[H]
\begin{center}
 \subfloat[]{\label{kappatheory}\includegraphics[width=0.50\textwidth]{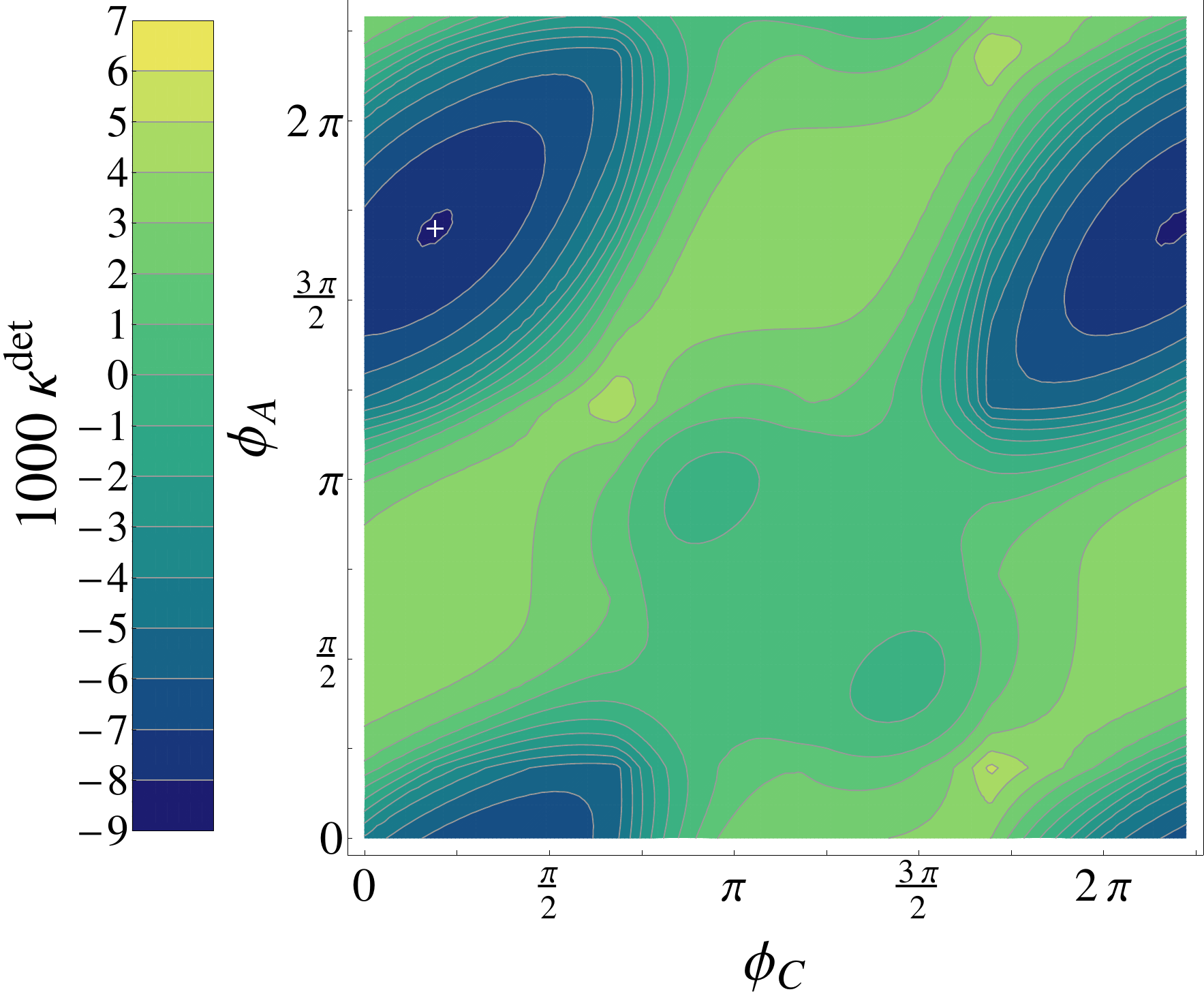}}  \ \ \
 \subfloat[]{\label{kappavsintensity}\includegraphics[width=0.435\textwidth]{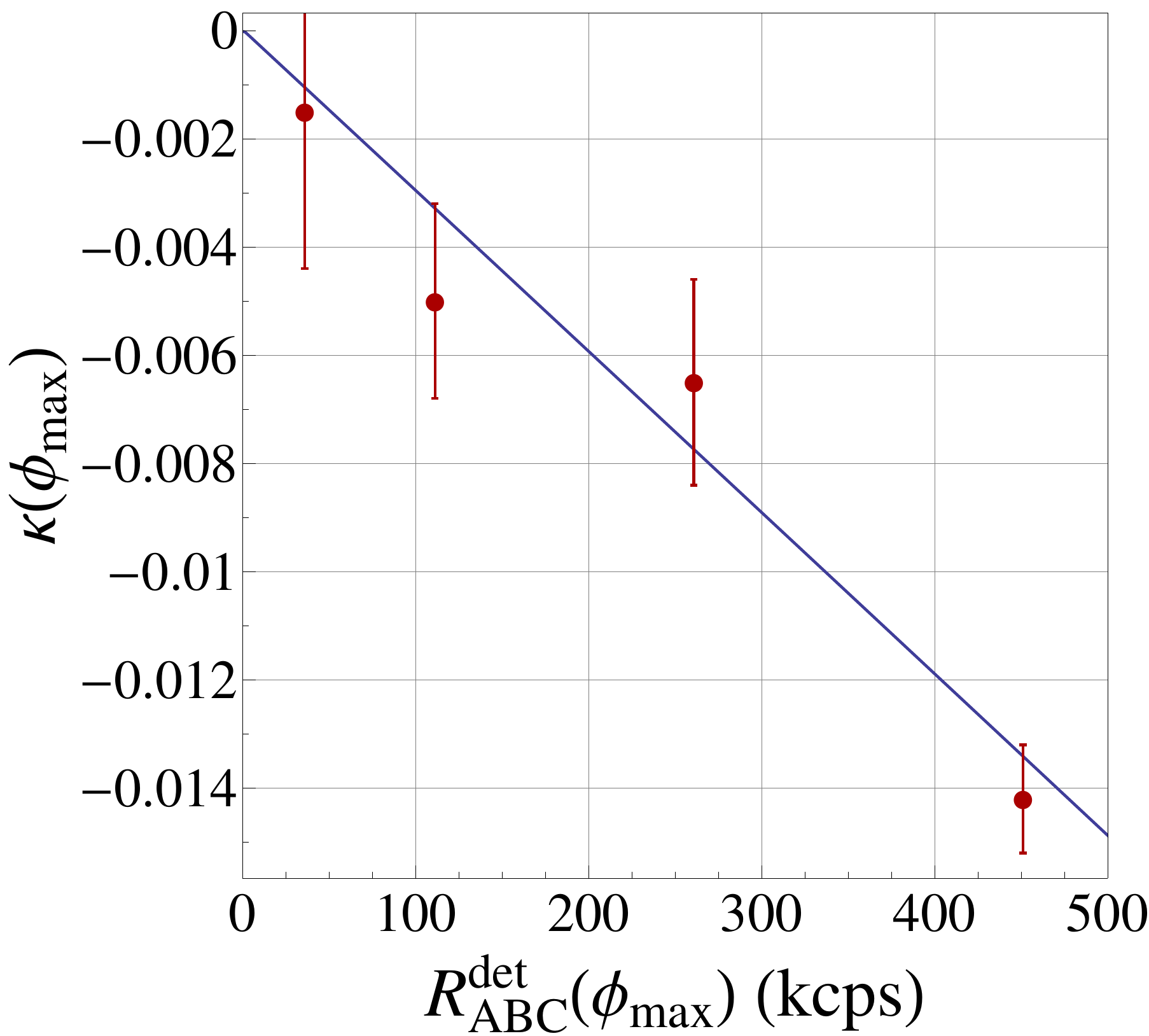}}
\end{center}
\caption{(a) Contour plot of the theoretical prediction of $\kappa^\mathrm{det}$ in phase space using the nonlinear model describing the detector using the independent single-path intensities as input parameters. The \textit{white cross} marks the position of the three-path maximum $\phi_\mathrm{max}$. In (b) one can see the relationship between the intensity at the three-path maximum, $R_{ABC}^\mathrm{det}(\phi_\mathrm{max})$, and the corresponding calculated $\kappa^\mathrm{det}(\phi_\mathrm{max})$ (\emph{solid line}) as well as the values $\kappa^\mathrm{exp}$ (\emph{circles}, for error bars see Table 1) measured at the same position.}
\label{kappatot}
\end{figure}

A good agreement between theory and experiment is observed, however some long time scale stability issues were found when different scans taken over several days were compared. To minimize the  influence of phase instability in the measurement we chose the three path maximum, which in Fig.~\ref{threepathdata} corresponds to $\phi_\mathrm{max}=(\phi_{C}, \phi_{A})=(0.19\pi, 1.7\pi)$ (marked by a cross), as the ideal position in phase space for more accurate $\kappa$ measurements. In Fig.~\ref{kappatheory} it becomes clear that this is one of the positions in phase space with the smallest gradient and it is also the position with the largest absolute deviation for $\kappa^\mathrm{det}$ due to detector nonlinearity. The hope is that other nonlinear effects would also be the largest at this position.

Each $\kappa$ measurement consisted of the average of one thousand runs of the eight necessary combinations at one fixed position in phase space. As mentioned above the three-path maximum was chosen as the preferred position in phase space. The phases corresponding to this position were found by first doing a fast scan of the entire two-dimensional space like the one shown in Fig.~\ref{measuredthreepath}. These angles were then used to take long $\kappa$ measurements at this preferred position. Here we were somewhat limited by the accuracy with which our motor allowed us to go to a predetermined position and ended up measuring for various count rates close to the phase space maximum as show in Table~\ref{kappatab}.

\begin{table}[tp]%
\centering%
\begin{tabular}{ ccccc }
$R_{ABC_0}^\mathrm{exp}$  & $\kappa^\mathrm{det}$ & $\kappa^\mathrm{exp}$ & $\Delta\kappa$ \\
\hline
$	35925	$&$	-0.0011	$&$	-0.0015	$&$	0.0029	$	\\	\hline
$	111288	$&$	-0.0033	$&$	-0.0050	$&$	0.0018	$	\\	\hline
$	260934	$&$	-0.0077	$&$	-0.0065	$&$	0.0019	$	\\	\hline
$	451121	$&$	-0.0134	$&$	-0.0142	$&$	0.0010	$	\\	\hline
\end{tabular}
\caption{Comparison of the theoretically predicted deviation from $\kappa=0$, using the detector nonlinearity model, $\kappa^\mathrm{det}$, and our actual measurement results, $\kappa^\mathrm{exp}$ for the position in phase space  given by $\phi_\mathrm{max}=(\phi_{C}, \phi_{A})=(0.19\pi, 1.7\pi)$, which is the intensity maximum of the three-path interference pattern, $R_{ABC}^\mathrm{det}$, as shown in Fig.~\ref{kappavsintensity}. Here, $\Delta\kappa$ is the standard error of the mean.}
\label{kappatab}
\end{table}

\section{Conclusions}

In conclusion our results show good agreement with our model including the detector nonlinearity. Future measurements will investigate different areas of Fig.~\ref{kappavsintensity} to see if the general predicted trend holds true. When phase stability in the interferometer is further improved it will also be of interest to investigate different areas of the phase space in Fig.~\ref{kappatheory}. We should also point out that the relationship shown in Fig.~\ref{kappavsintensity} depends not only on the dead time of the detector but also on the photon statistics of the light source used. The ideal case would be the emission from a single quantum emitter with the excitation events separated by more than a detector dead time. In this scenario no effects due to detector nonlinearity would be seen and the detector dead time would only limit the maximum achievable count rate.

Furthermore, with this experiment we can bound the absolute amount of multi-order interference as measured by $|\kappa|$ to be smaller than $0.0015 \pm	0.0029$. This represents an improvement of about one order of magnitude over the best previous limit \cite{Sinha10}. For this new limit we have not used any correction of systematic errors such as the detector nonlinearity model. With a more accurate exploration of the relation between count rate and $\kappa$ we expect to be able to reduce the bound by several orders of magnitude using the same experimental set-up.

\bibliographystyle{spphys}
\bibliography{threepath}
\end{document}